\documentclass[a4paper,prl,twocolumn,amsmath,amssymb,longbibliography]{revtex4-1}
\usepackage{bm}
\usepackage{graphicx,color,soul}
\usepackage[colorlinks,linkcolor=magenta,citecolor=magenta]{hyperref}
\usepackage[ruled,vlined,norelsize]{algorithm2e}

\newcommand{\dd}{\mathrm{d}}
\newcommand{\td}[2]{\frac{\dd #1}{\dd #2}}
\newcommand{\pd}[2]{\frac{\partial #1}{\partial #2}}

\newcommand{\mean}[1]{\langle #1 \rangle}
\newcommand{\Int}[1]{\int\dd #1\;}
\newcommand{\IInt}[3]{\int_{#2}^{#3}\dd #1\;}
\renewcommand{\vec}[1]{\mathbf #1}

\newcommand{\kap}{\kappa}
\newcommand{\lam}{\lambda}
\newcommand{\vhi}{\varphi}
\newcommand{\sig}{\sigma}

\newcommand{\kT}{k_\text{B}T}
\newcommand{\tx}{\tau_\text{r}}
\newcommand{\Dc}{D_\text{c}}
\newcommand{\x}{\vec r}
\newcommand{\nois}{\bm\xi}

\begin{document}

\title{Efficiency of isothermal active matter engines: Strong driving beats weak driving}

\author{Thomas Speck}
\affiliation{Institut f\"ur Physik, Johannes Gutenberg-Universit\"at Mainz, Staudingerweg 7-9, 55128 Mainz, Germany}


\begin{abstract}
  We study microscopic engines that use a single active particle as their ``working medium''. Part of the energy required to drive the directed motion of the particle can be recovered as work, even at constant temperature. A wide class of synthetic active particles can be captured by schematically accounting for the chemical degrees of freedom that power the directed motion without having to resolve the exact microscopic mechanism. We derive analytical results for the quasi-static thermodynamic efficiency, i.e., the fraction of available chemical energy that can be recovered as mechanical work. While this efficiency is vanishingly small for colloidal particles, it increases as the dissipation is increased beyond the linear response regime and goes through a maximum at large propulsion speeds. Our results demonstrate that driving beyond the linear response regime has non-trivial consequences for the efficiency of active engines.
\end{abstract}

\maketitle


Macroscopic engines that convert heat into useable work have been an important factor driving the industrial revolution and the development of thermodynamics in the 18th and 19th century~\cite{muller}. These engines operate cyclically between two (or more) heat baths, with a working medium taking up heat from the hotter and dumping it into the colder heat bath, the temperatures $T_\text{h,c}$ of which limit the efficiency $\eta\leqslant\eta_\text{C}$ to the famous Carnot efficiency $\eta_\text{C}=1-T_\text{c}/T_\text{h}$. More recently, the understanding of microscopic engines that operate in the presence of strong thermal fluctuations has gained interest. In the extreme limit, the working medium can be reduced to a single particle~\cite{martinez17}, which has been demonstrated experimentally for a trapped colloidal particle~\cite{blickle11,martinez15} and a trapped single ion~\cite{rossnagel16} (with the perspective to exploit genuine quantum effects~\cite{gelbwaser18}).

The second law of thermodynamics prevents the conversion of heat from a single \emph{equilibrium} heat bath into work without dumping some heat back into a colder heat bath. On the microscale, changing the temperature is difficult and often undesired. Cyclic operation is required to return the working medium to its initial state before the next cycle begins. In contrast, the molecular machines operating living matter cycle through several molecular conformations by converting chemical (free) energy at constant temperature, typically without a cyclic variation of parameters. The (stochastic) thermodynamics of Brownian motors has been studied extensively~\cite{julicher97,parrondo02,kolomeisky07,seifert11a,zimm12}. Similar in spirit to Brownian motors are colloidal engines--active particles--that convert chemical energy into directed motion through a viscous environment~\cite{colberg14,bech16}. The unavoidable rotational fluctuations randomize this motion on long time scales. Still, the directed motion can be exploited to extract work, \emph{e.g.}, through transporting cargo~\cite{baraban12,niu18} and harvesting the forces on embedded obstacles~\cite{leonardo10,sokolov10,pietzonka19}. Recently, isothermal cyclic engines that extract work from a single heat reservoir through employing an active fluid as working medium have been realized with bacteria~\cite{krishna16} and explored further theoretically~\cite{zakine17,holubec20,ekeh20,kumari20,malgaretti21}. Current attempts to build a thermodynamic framework for active fluids focus on observable degrees of freedom and neglect the contribution of (chemical) degrees of freedom underlying self-propulsion~\cite{shankar18,dabelow19,szamel19,fodor20}, or are restricted to the linear response regime~\cite{huang19,gaspard19,markovich21}. However, for possible applications it is imperative to understand the full efficiency including the energy budget to maintain the working medium away from equilibrium.

\begin{figure}[b!]
  \centering
  \includegraphics{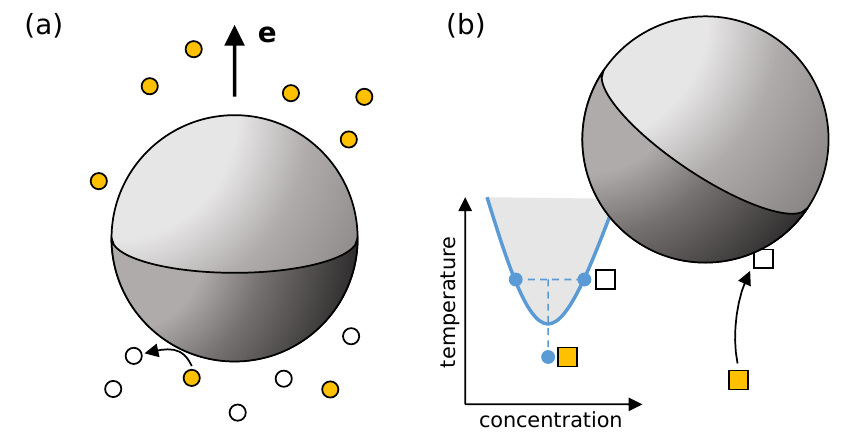}
  \caption{Hidden chemical degrees of freedom drive self-propulsion. (a)~Diffusiophoretic Janus particle propelled through the conversion $\bullet\to\circ$ of (neutral) molecular solutes on its catalytic hemisphere (dark)~\cite{gole05}. The conversion is driven by the difference $\Delta\mu=\mu_\bullet-\mu_\circ$ of chemical potential. The gradient of solutes inside the hydrodynamic boundary layer induces a stress and thus flow of the solvent~\cite{sabass12}, with the particle moving along its unit orientation $\vec e$. (b)~Janus particle moving in a binary solvent (water+molecular solute) close to its lower critical point~\cite{samin15}. The dark hemisphere is locally heated through light absorption, quenching the solvent to the two-phase region (inset). Bringing a homogeneous solvent parcel (filled box) close to the heated surface converts the box to solute-rich (or solute-poor, depending on the surface properties), which liberates the free energy $\Delta\mu$.}
  \label{fig:sketch}
\end{figure}

To fill this gap, here we consider a single active particle as working medium. We make one crucial assumption, namely the \emph{tight coupling} between (schematic) chemical events and the translation of the colloidal particle, which excludes more complex organisms like swimming bacteria. This assumption allows to infer the dissipation without having to resolve the exact microscopic mechanism responsible for the directed motion. Exploiting the local detailed balance condition restricting the rates for the chemical events, we calculate the \emph{thermodynamic} efficiency in the relevant limit that the step size is much smaller than the particle size. This efficiency is vanishingly small, and most of the available (free) energy is dissipated to drag along the solvent. Our approach covers a wide range of phoretic mechanisms reported in the literature, two of which are sketched in Fig.~\ref{fig:sketch}. But even if the microscopic dynamics involved idle cycles (dissipation without directed motion), our results still serve as an upper bound to the thermodynamic efficiency of cyclic active engines.

To be specific, the active particle moves in two dimensions and is driven by the schematic conversion of substrate into product. We do not resolve the exact mechanism (Fig.~\ref{fig:sketch} shows two possible experimental realizations) but assume that each conversion liberates the free energy $\Delta\mu$ and translates the particle by a (small) distance $\lam$ along its unit orientation $\vec e=(\cos\vhi,\sin\vhi)^T$~\cite{speck18,fischer19}. Throughout, we employ dimensionless quantities measuring time in units of the orientational correlation time $\tx$, lengths in units of $\ell=(\kT\mu_0\tx)^{1/2}$ with bare translational mobility $\mu_0$, and energies in units of the thermal energy $\kT$. If rotational and translational diffusion are coupled due to the no-slip boundary condition for the surrounding solvent then $\ell=\sig_\text{H}/\sqrt 3$ with hydrodynamic diameter $\sig_\text{H}$ of the active particle.


We start by considering a free particle with rates $\kap_0^\pm=\kap_0e^{\pm\Delta\mu/2}$ for the chemical events. For tight coupling as assumed here, the bare propulsion speed (averaged over the chemical fluctuations) reads
\begin{equation}
  v_0 = \lam(\kap_0^+-\kap_0^-) = 2\lam\kap_0\sinh\left(\frac{\Delta\mu}{2}\right).
  \label{eq:v0}
\end{equation} 
The average injected power $P_\text{in,0}=\Delta\mu v_0/\lam$ is simply the amount of free energy $\Delta\mu$ liberated in each chemical event times the average net number of events per time, $\dot n=\kap_0^+-\kap_0^-=v_0/\lam$. This chemical work changes the free energy of the chemical reservoirs with the excess dissipated as heat~\cite{seifert11a}.

To estimate the magnitudes of the parameters $\lam$ and $\kap_0$, we consider colloidal Janus particles that are driven by the reversible demixing of a binary solvent due to local heating~\cite{butt13,samin15,solano17,schmidt19}, cf. Fig.~\ref{fig:sketch}(b). For example, the particles used in Ref.~\citenum{butt13} had a diameter $\sig_\text{H}\simeq 4.3\,\mu$m and reached speeds of order $1\,\mu$m/s, which together with $\tx\simeq200\,$s leads to $v_0\sim80$. For $\Delta\mu$ of order unity this implies $\lam\kap_0\sim10$ in Eq.~\eqref{eq:v0}. Turning to more explicit models of diffusiophoretic particles~\cite{sabass12} shows that the displacement $\lam\ll 1$ is related to the square of the thickness of the boundary interaction layer and thus very small (compared to the size of the particle), which is compensated by a large attempt rate $\kap_0\gg 1$ so that the product $\lam\kap_0$ is of order unity. Note that all three parameters $\Delta\mu$, $\kap_0$, and $\lam$ are influenced by the specific propulsion mechanism and are neither constant nor independent.

\begin{figure}[t]
  \centering
  \includegraphics{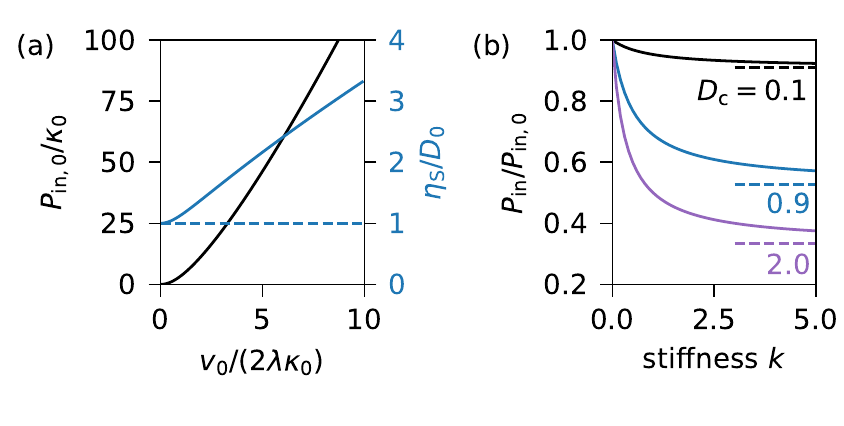}
  \caption{(a)~Left axis: dissipated power $P_\text{in,0}$ to propel a free active particle as a function of the reduced speed $v_0/(2\lam\kap_0)$. Right axis: Stokes efficiency $\eta_\text{S}/D_0$ (with respect to the power needed to drag a passive particle with the same speed). The dashed line is the limiting efficiency for weak driving set by the fluctuations of the chemical events $D_0=\lam^2\kap_0$. (b)~Dissipated power [Eq.~\eqref{eq:q}] divided by $P_\text{in,0}$ for an active particle moving in a harmonic trap. Dashed lines indicate the limiting values $P_\text{in}/P_\text{in,0}\to(1+\Dc)^{-1}$ for very stiff traps.}
  \label{fig:diss}
\end{figure}

As a first measure of efficiency, we consider the Stokes efficiency $\eta_\text{S}=v_0^2/P_\text{in,0}$ which compares the injected power to the power $v_0^2$ necessary to move a passive bead with the propulsion speed $v_0$ against the viscous drag~\cite{wang02,sabass12,zimm12}. Note that $\eta_\text{S}$ is not the thermodynamic efficiency, and not bounded by one. Still, $\eta_\text{S}$ provides a useful measure how much of the available chemical energy is actually converted to directed motion. Plugging in the power $P_\text{in,0}=\Delta\mu v_0/\lam$, we find
\begin{equation}
  \eta_\text{S}=D_0\hat v/\sinh^{-1}\hat v
  \label{eq:eta:s}
\end{equation}
as a function of the reduced speed $\hat v=v_0/(2\lam\kap_0)$ with $D_0=\lam^2\kap_0\sim\lam\ll 1$ showing that only a small fraction of the injected power is converted into moving the particle against the viscous drag. The power and Stokes efficiency are plotted in Fig.~\ref{fig:diss}(a), and show that strong driving increases the efficiency, \emph{e.g.} at a speed $\hat v=10$ the efficiency is increased by a factor of more than three compared with a weakly driven active particle.

How does the increase of the Stokes efficiency relate to the efficiency of active engines? To address this question, we calculate the thermodynamic efficiency for our active particle in an optical trap with potential energy $U(\x)=\tfrac{1}{2}k\x^2$ that is operated cyclically. As forward and backward rates for the chemical events we choose
\begin{equation}
  \kap^\pm(\x,\vhi)
  = \kap_0e^{-k\lam^2/2}e^{\pm(\Delta\mu-k\lam\vec e\cdot\x)/2},
\end{equation}
which obey the local detailed balance condition
\begin{equation}
  \frac{\kap^+(\x)}{\kap^-(\x+\lam\vec e)} = e^{\Delta\mu-[U(\x+\lam\vec e)-U(\x)]}.
  \label{eq:ldb}
\end{equation}
This condition ensures that the active particle coupled to the two chemical reservoirs obeys both the first and second law along single stochastic trajectories~\cite{seifert12,sm}.

The average injected power now becomes
\begin{equation}
  P_\text{in} = \frac{\Delta\mu\mean{v}}{\lam}
  \label{eq:q:def}
\end{equation}
with propulsion speed $v(\x,\vhi)=\lam(\kap^+-\kap^-)$ that depends on the position and orientation of the particle. In the following, we exploit the smallness of $\lam\ll 1$ and expand to second order,
\begin{equation}
  v(\x,\vhi) \approx v_0-k\Dc\vec e\cdot\x,
  \label{eq:v:lin}
\end{equation}
with bare speed $v_0$ [Eq.~\eqref{eq:v0}]. Here,
\begin{equation}
  \Dc = \lam^2\kap_0\cosh\left(\frac{\Delta\mu}{2}\right) = D_0\sqrt{1+\hat v^2}
\end{equation}
is the contribution to the translational diffusion coefficient due to the chemical reactions. To order $\lam^2$ it remains state independent and for small propulsion speeds $\Dc\approx D_0$. The second term in Eq.~\eqref{eq:v:lin} is the force projected onto the orientation so that the particle speeds up if its orientation points towards the origin and slows down if pointing outward.


The expression Eq.~\eqref{eq:q:def} is well-known from the study of molecular motors~\cite{julicher97,parrondo02,kolomeisky07} but differs fundamentally from attempts to identify dissipation and entropy production of active particles from their equations of motion ignoring the (chemical) degrees of freedom underlying self-propulsion~\cite{shankar18,dabelow19,szamel19,holubec20}. For example, adopting the perspective that a thermostat with noise $\nois$ injects the average power $P_\text{T}=\mean{\dot\x\cdot(\dot\x-\nois)}$ to keep the system in the steady state~\cite{ekeh20,fodor21}, together with the stochastic equation of motion $\dot\x=v_0\vec e-k\x+\nois$ one obtains
\begin{equation}
  P_\text{T} = v_0^2 - v_0k\mean{\vec e\cdot\x}.
\end{equation}
On the other hand, expanding Eq.~\eqref{eq:v0} to linear order of $\Delta\mu$ yields $v_0\approx\lam\kap_0\Delta\mu$ and plugging this together with Eq.~\eqref{eq:v:lin} into Eq.~\eqref{eq:q:def}, we arrive at the injected power
\begin{equation}
  P_\text{in,lin} \approx \frac{v_0^2}{D_0} - v_0k\mean{\vec e\cdot\x} \neq P_\text{T}
  \label{eq:q:lin}.
\end{equation}
This result for $P_\text{in,lin}$ has also been obtained by Pietzonka and Seifert for a coarse-grained lattice model~\cite{pietzonka17}. Only for $D_0=1$ would this expression coincide with $P_\text{T}$, but in fact $D_0\sim\lam\ll 1$ as argued above. More importantly, for the model considered here Eq.~\eqref{eq:q:lin} only holds in the \emph{linear regime} for small $\Delta\mu$. In the following, we explore the consequences of strong driving with $\Delta\mu>1$.


To calculate the average input power for motion confined by the harmonic trap, we require the cross correlations $\chi_1=\mean{\vec e\cdot\x}=\Int{^2\x\dd\vhi}(\vec e\cdot\x)\psi$ with joint probability $\psi(\x,\vec e;t)$ of position $\x$ and unit orientation $\vec e$ obeying the evolution equation $\partial_t\psi=\mathcal L_0\psi+\mathcal L_1\psi$. Here,
\begin{equation}
  \mathcal L_0\psi = k\nabla\cdot(\x\psi) + \nabla^2\psi + \partial_\vhi^2\psi
  \label{eq:L0}
\end{equation}
describes the passive translational and rotational diffusion. The discrete displacements of the particle along its orientation can be modeled by the master equation (only arguments different from $\x$ are indicated)~\cite{speck18}
\begin{multline}
  \mathcal L_1\psi = \kap^+(\x-\lam\vec e)\psi(\x-\lam\vec e) + \kap^-(\x+\lam\vec e)\psi(\x+\lam\vec e) \\ - (\kap^++\kap^-)\psi.
  \label{eq:L1}
\end{multline}
We then find $\partial_t\chi_1=-(1+k)\chi_1+\mean{v}$, where the second term follows from inserting Eq.~\eqref{eq:L1}, shifting arguments, and $\vec e\cdot(\x\pm\lam\vec e)=\vec e\cdot\x\pm\lam$~\cite{sm}. The average propulsion speed Eq.~\eqref{eq:v:lin} can now be written $\mean{v}\approx v_0-k\Dc\chi_1$. Setting the time derivative to zero in the steady state, we plug the result
\begin{equation}
  \chi_1 = \frac{v_0}{1+k(1+\Dc)}
  \label{eq:chi1}
\end{equation}
into Eq.~\eqref{eq:q:def} to obtain
\begin{equation}
  P_\text{in} = P_\text{in,0}\left[1-\frac{k\Dc}{1+k(1+\Dc)}\right].
  \label{eq:q}
\end{equation}
In Fig.~\ref{fig:diss}(b), we show that the reduced injected power $P_\text{in}/P_\text{in,0}\leqslant 1$ decreases when increasing the trap strength $k$ since the confining potential induces ``backsteps'' that restore product and thus reduce the dissipation.


\begin{figure}[t]
  \centering
  \includegraphics{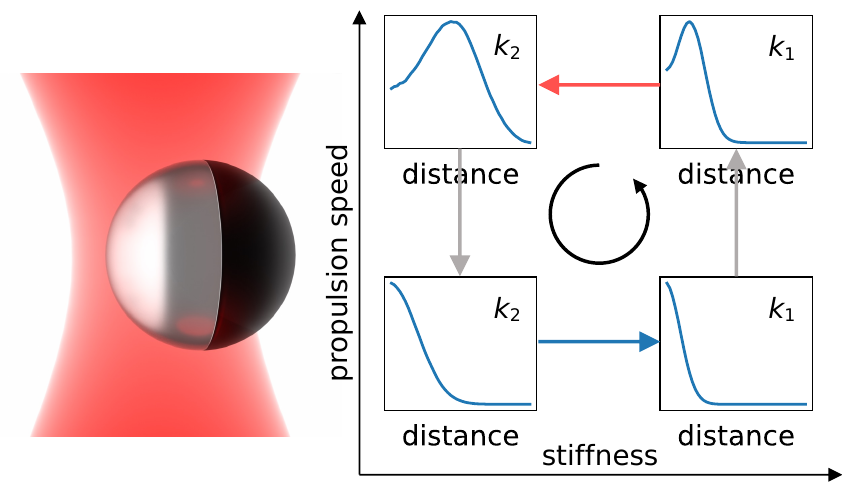}
  \caption{Cyclic isothermal active engine: a self-propelled Janus particle is trapped by optical tweezers imposing a harmonic potential with stiffness $k$. A Stirling-like cycle can be realized at constant temperature through changing propulsion speed and trap stiffness. Red arrow: expansion at high speed delivering work against the trap. Blue arrow: compression at low speed requiring work. Gray vertical arrows: isochoric changes of propulsion speed. The histograms show numerical results [solving the dynamics corresponding to Eqs.~\eqref{eq:L0} and \eqref{eq:L1}] for the stationary probability to find the particle a given distance from the origin (parameters: $k_1=3$, $k_2=1$ and $\Delta\mu_1=1$, $\Delta\mu_2=3$ with $\lam=10^{-3}$ and $\lam\kap_0=1$). Note the departure from the Gaussian distribution~\cite{poto12}.}
  \label{fig:cyc}
\end{figure}

We now move on to cyclic isothermal engines with the single active particle as its working medium (in the spirit of Refs.~\citenum{blickle11,martinez15}). There are two processes, changing the trap strength $k$ at constant bare propulsion speed $v_0$ and changing $v_0$ at constant $k$ (Fig.~\ref{fig:cyc}). We consider the quasi-static limit so that the distribution $\psi(\x,\vhi)$ depends on time only through the instantaneous values of $k$ and $v_0$. Changing the trap stiffness from $k_1$ to $k_2$ during the time $\tau/2$ at constant speed $v_0$, the work due to the injected power becomes
\begin{equation}
  W_\text{in}^\leftarrow = \IInt{t}{0}{\tau/2}P_\text{in} = \frac{\tau/2}{\Delta k}\IInt{k}{k_1}{k_2}P_\text{in} = \frac{\tau}{2}P_\text{in,0}f
\end{equation}
with $\Delta k=k_2-k_1$ and function
\begin{multline}
  f(k_1,k_2,\Dc) = \\ \frac{1}{1+\Dc} + \frac{\Dc}{(1+\Dc)^2\Delta k}\ln\frac{1+k_2(1+\Dc)}{1+k_1(1+\Dc)},
\end{multline}
which is symmetric with respect to exchanging $k_1\leftrightarrow k_2$. Changing $k$ also implies that the work
\begin{equation}
  W_\text{out}^\leftarrow = \IInt{k}{k_1}{k_2} \left\langle\pd{U}{k}\right\rangle = \frac{1}{2}\IInt{k}{k_1}{k_2}\mean{\x^2}
  \label{eq:w:out}
\end{equation}
is performed against the potential energy~\cite{blickle11,speck11,sm}. The second moment $\mean{\x^2}$ can be calculated along the same lines as for the cross-correlations $\chi_1$, leading to
\begin{align}
  \nonumber
  \partial_t\mean{\x^2} &= \Int{^2\x\dd\vhi} \x^2(\mathcal L_0\psi+\mathcal L_1\psi) \\ &= -2k\mean{\x^2} + 4 + 2\mean{v\vec e\cdot\x} + 2\Dc.
\end{align}
To proceed, we insert the expansion Eq.~\eqref{eq:v:lin} up to order $\lam^2$ and, in the steady state, we set the time derivative to zero to obtain
\begin{equation}
  k\mean{\x^2} = 2 + v_0\chi_1 + (1-k\chi_2)\Dc.
  \label{eq:r2}
\end{equation}
While $\chi_2=\mean{(\vec e\cdot\x)^2}$ can be calculated analytically~\cite{sm}, for now we focus on the leading contribution
\begin{align}
  \nonumber
  W_\text{out} &= W_\text{out}^\rightarrow(v_1) + W_\text{out}^\leftarrow(v_2) \\
  &= \frac{1}{2}(v_2^2-v_1^2)\ln\left(\frac{k_2}{k_1}\frac{1+k_1}{1+k_2}\right) + \mathcal O(\Dc)
\end{align}
to the work during one cycle assuming $\Dc\ll 1$. The work $W_\text{out}$ becomes negative, and thus in principle available, for $v_2>v_1$ and $k_2<k_1$ leading to the cycle sketched in Fig.~\ref{fig:cyc}, where the trap is compressed for lower speed $v_1$ and expanded at higher speed $v_2$.


The thermodynamic efficiency $\eta=-W_\text{out}/W_\text{in}\leqslant 1$ is strictly bounded by one, which here is guaranteed by the rates $\kap^\pm$ obeying the local detailed balance condition~\eqref{eq:ldb}~\cite{sm}. The work performed by the reservoirs is $W_\text{in}=W_\text{in}^\rightarrow(v_1)+W_\text{in}^\leftarrow(v_2)$ neglecting the work to switch the propulsion speed. We now derive an expression for the efficiency for large speeds but so that $\Dc=D_0\sqrt{1+\hat v^2}\ll 1$ remains small, and we consider only the leading contribution in an expansion of $\Dc$. First, for small speeds in the linear regime $W_\text{in,lin}=(\tau/D_0)\tfrac{1}{2}(v_2^2+v_1^2)f(k_1,k_2,D_0)$ and thus
\begin{equation}
  \eta_\text{lin} = \frac{-W_\text{out}}{W_\text{in,lin}} = \frac{D_0}{\tau}\frac{v_2^2-v_1^2}{v_2^2+v_1^2}\ln\left(\frac{k_1}{k_2}\frac{1+k_2}{1+k_1}\right)
\end{equation}
with $f=1+\mathcal O(D_0)$. It is easy to see that this efficiency is maximized by performing the compression of the trap with a passive particle, $v_1=0$. The linear efficiency then is independent of the speed, $\eta_\text{lin}\propto D_0/\tau$, and its magnitude is set by $D_0$ as for the Stokes efficiency. Going beyond the linear regime (still to leading order of $\Dc$), we find
\begin{equation}
  \frac{\eta}{\eta_\text{lin}} = \frac{W_\text{in,lin}}{W_\text{in}} = \frac{\hat v_2^2+\hat v_1^2}{\hat v_2\sinh^{-1}\hat v_2+\hat v_1\sinh^{-1}\hat v_1} \geqslant 1
\end{equation}
with reduced speeds $\hat v_i=v_i/(2\lam\kap_0)$. The optimum is still to perform the compression with a passive particle, whence the relative efficiency $\eta/\eta_\text{lin}=\hat v_2/\sinh^{-1}\hat v_2=\eta_\text{S}/D_0$ becomes equal to the relative Stokes efficiency for a free particle [Eq.~\eqref{eq:eta:s}]. In particular, increasing the (reduced) speed $\hat v_2$ increases the efficiency considerably but also the extracted work $W_\text{out}\sim v_2^2$.


\begin{figure}[t]
  \centering
  \includegraphics{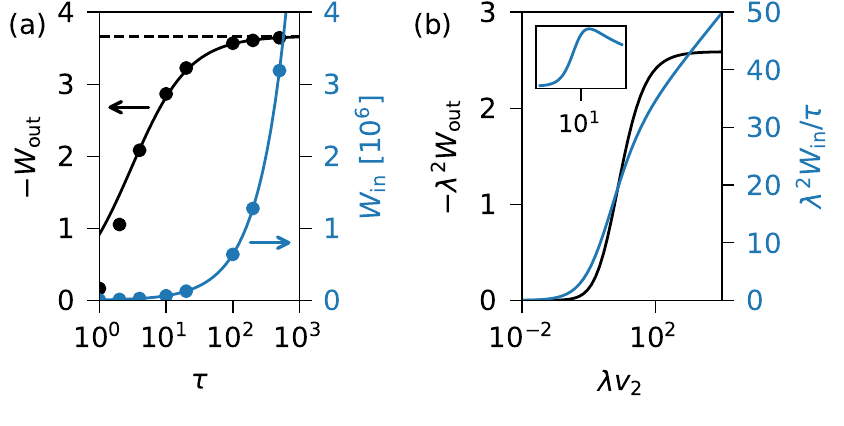}
  \caption{(a)~Numerical results for finite cycle times $\tau$ (parameters as in Fig.~\ref{fig:cyc} but with passive compression, $v_1=0$). Left axis shows extracted work $-W_\text{out}^\tau$ (symbols) with the dashed line indicating the quasi-static limit $-W_\text{out}$. The solid line is $-W_\text{out}\tau/(\tau_0+\tau)$ with $\tau_0=3$. Right axis shows the injected work $W_\text{in}=\mean{n}\Delta\mu$ (symbols). The solid line is $\bar P_\text{in}\tau$ with average injected power $\bar P_\text{in}\simeq6.4\times10^3$. (b)~Quasi-static work and injected power from numerical integration as a function of speed $v_2$. Quantities are scaled so that the magnitudes do not depend on $\lam$. The inset shows the efficiency $\eta=-W_\text{out}/W_\text{in}$ as a function of $\lam v_2$.}
  \label{fig:num}
\end{figure}

In Fig.~\ref{fig:num}(a), we plot the average extracted work $-W_\text{out}^\tau$ as a function of the cycle time $\tau$ solving the dynamics corresponding to Eqs.~\eqref{eq:L0} and \eqref{eq:L1} numerically (same parameters as for Fig.~\ref{fig:cyc} but with $v_1=0$). We see that the extracted work is a few thermal energies and converges to the quasi-static value $W_\text{out}\simeq-3.67$ obtained from numerically integrating Eq.~\eqref{eq:w:out} together with Eq.~\eqref{eq:r2}. In the simulations, we have direct access to the net number of chemical events $n$ and thus $W_\text{in}=\mean{n}\Delta\mu$, which is also plotted in Fig.~\ref{fig:num}(a). The magnitude of the injected work is $10^6$ times larger, and consequently $\eta\sim 10^{-6}$.

It is instructive to consider the limit of large speeds, although those might not be realizable with experimental active particles. While we still assume $\lam\ll 1$, the coefficient $\Dc\approx\lam v_0/2$ is no longer small and we take into account the full solution for $\mean{\x^2}$, which reaches a plateau $\mean{\x^2}\to4/(\lam k)^2$~\cite{sm}. Fig.~\ref{fig:num}(b) shows for $\lam\kap_0=1$ the result of numerically integrating Eq.~\eqref{eq:w:out}. The extracted work saturates for large speeds, which follows the behavior of the second moment $\mean{\x^2}$. The injected work $W_\text{in}$ continues to rise as the speed is increased, which implies that the efficiency goes through a maximal value for large but finite speed [inset Fig.~\ref{fig:num}(b)]. The position $v_0^\ast\sim 10/\lam\gg 10$ of this maximum depends on $\lam$ implying a large speed for $\lam\ll1$.


To conclude, we have shown that strongly driven active particles, \emph{i.e.} large affinities $\Delta\mu$, can improve the efficiency of active engines by several hundred percent [cf. Fig.~\ref{fig:diss}(a)]. While one might expect the maximal efficiency to be realized in the linear response regime, some models for molecular motors also show that far from equilibrium the efficiency can increase with $\Delta\mu$~\cite{parmeggiani99}. The total efficiency depends strongly on how efficient an active particle converts the chemical energy into directed motion. General arguments indicate that this efficiency is limited by the thickness of the boundary layer with respect to the particle size~\cite{sabass10}, and thus is small for $\mu$m-sized particles. Note that we did not consider the \emph{total} entropy production [\emph{e.g.}, in Fig.~\ref{fig:sketch}(b) there is clearly another ``housekeeping'' heat current to maintain the temperature profile] but the (chemical) energy that is in principle available for motion. Still, active colloidal engines can only access a tiny portion of this energy, and previous estimates of efficiencies for active engines are far too optimistic~\cite{ekeh20,holubec20}. Comparably small efficiencies have also been reported for a colloidal clutch~\cite{williams15}. In contrast, molecular motors such as F$_1$-ATPase can operate close to the efficiency limit~\cite{seifert11,zimm12}. The reason is an inherent physical limitation for motion in the Stokes regime: larger particles have to ``drag'' more solvent. In this respect, due to their size catalytic enzymes~\cite{sengupta13,jee18,ghosh21} might become an interesting class of active matter and potential building blocks for the extraction of work. Our conceptual results illustrate the importance and potential of the non-linear behavior of dissipation for the design of engines on the micro and nanoscale.


\begin{acknowledgments}
  I thank Udo Seifert for inspiring and illuminating discussions. William Janke is acknowledged for rendering the trapped particle in Fig.~\ref{fig:cyc}.
\end{acknowledgments}


%

\newpage

\section{Cross correlations}

The dynamics of the single particle is determined by the two Fokker-Planck operators
\begin{equation}
  \mathcal L_0\psi = k\nabla\cdot(\x\psi) + \nabla^2\psi + \partial_\vhi^2\psi
  \label{eq:L0}
\end{equation}
and
\begin{multline}
  \mathcal L_1\psi = \kap^+(\x-\lam\vec e)\psi(\x-\lam\vec e) + \kap^-(\x+\lam\vec e)\psi(\x+\lam\vec e) \\ - [\kap^+(\x)+\kap^-(\x)]\psi(\x).
  \label{eq:L1}
\end{multline}
We calculate a closed evolution equation for the cross correlations $\chi_1(\x,\vhi;t)=\Int{^2\x\dd\vhi}(\vec e\cdot\x)\psi(\x,\vhi;t)$. The time derivative reads
\begin{equation*}
  \partial_t\chi_1 = \Int{^2\x\dd\vhi} (\vec e\cdot\x)(\mathcal L_0\psi+\mathcal L_1\psi).
\end{equation*}
We first insert $\mathcal L_0$ and perform integrations by parts
\begin{multline*}
  \Int{^2\x\dd\vhi} [-\nabla(\vec e\cdot\x)]\cdot[k\x\psi+\nabla\psi] + \Int{^2\x\dd\vhi}[\partial_\vhi^2(\vec e\cdot\x)]\psi \\ = -k\chi_1 - \chi_1
\end{multline*}
with vanishing boundary terms. For $\mathcal L_1$ let us consider
\begin{multline*}
  \Int{^2\x\dd\vhi} (\vec e\cdot\x)\kap^+(\x-\lam\vec e)\psi(\x-\lam\vec e) \\ = \Int{^2\x\dd\vhi} [\vec e\cdot(\x+\lam\vec e)]\kap^+(\x)\psi(\x)
\end{multline*}
shifting the integration variable $\x\to\x+\lam\vec e$ with unit Jacobian. The first term cancels with the diagonal part of the operator and the second term yields $\lam\kap^+$. The same calculation holds for $\kap^-$ but now shifting $\x\to\x-\lam\vec e$, and thus we obtain $\lam(\kap^+-\kap^-)=v$ leading to the exact result
\begin{equation*}
  \partial_t\chi_1 = -(1+k)\chi_1 + \mean{v}.
\end{equation*}
Inserting the expansion $\mean{v}\approx v_0-k\Dc\chi_1$ leads to the result given in the main text.

For $\chi_2=\mean{(\vec e\cdot\x)^2}$ we first consider the more general expression $\chi'=\mean{(a_{ij}e_ix_j)^2}$ leading to
\begin{multline*}
  \partial_t\chi' = -2(1+k)\chi' + 2a_{ik}a_{jk}\mean{e_ie_j} + 2\left(a_{ij}\pd{e_i}{\vhi}x_j\right)^2 \\
  + 2(a_{ij}e_ie_j)\mean{va_{ij}e_ix_j} + 2(a_{ij}e_ie_j)^2\mean{\Dc}.
\end{multline*}
The definition of $\chi_2$ corresponds to $a_{ij}=\delta_{ij}$. Using that $\partial_\vhi e_1=-e_2$ and $\partial_\vhi e_2=e_1$ we define $\chi_2^\perp$ for $a_{11}=a_{22}=0$ and $a_{12}=-a_{21}=1$ (the Levi-Civita symbol). In the steady state, the result can be gathered into the matrix equation
\begin{equation*}
  \left(\begin{array}{cc}
    1 + k(1+\Dc) & -1 \\ 1 & 1+k
  \end{array}\right)\left(\begin{array}{c}
    \chi_2 \\ \chi_2^\perp
  \end{array}\right) = \left(\begin{array}{c}
    1 + \Dc + v_0\chi_1 \\ 1
  \end{array}\right)
\end{equation*}
with solution
\begin{equation}
  \chi_2 = \frac{1+(1+k)(D+v_0\chi_1)}{2+k+Dk(1+k)}
  \label{eq:chi2}
\end{equation}
where $D=1+\Dc$.

\section{Large speeds}

\begin{figure}[t]
  \centering
  \includegraphics{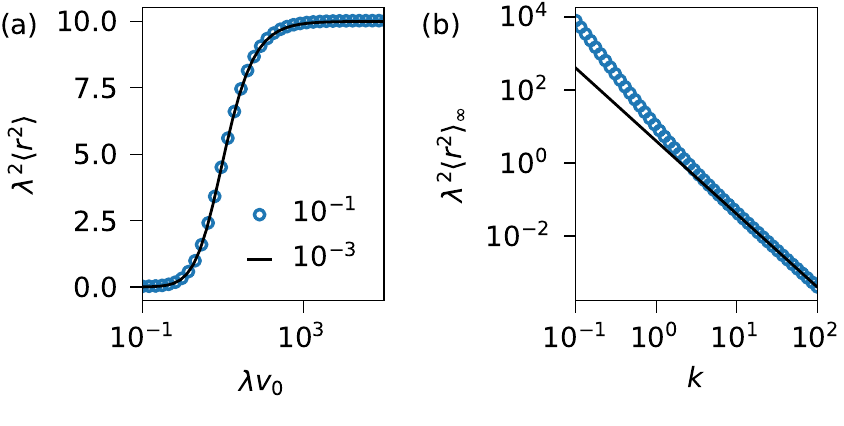}
  \caption{(a)~Second moment as a function of speed for two values $\lam=10^{-1},10^{-3}$. Axes are rescaled so that the function is independent of $\lam$. (b)~The limiting plateau value $\mean{\x^2}_\infty$ as a function of $k$. The solid line shows $4/(\lam k)^2$.}
  \label{fig:limit}
\end{figure}

We now consider large speeds with $\Dc\approx\tfrac{\lam}{2}v_0\gg 1$ after inserting $\hat v=v_0/(2\lam\kap_0)$. We expand the cross correlations to leading order,
\begin{equation*}
  \chi_1 = \frac{v_0}{1+k(1+\Dc)} \approx \frac{2}{\lam k} - 4\frac{1+k}{(\lam k)^2v_0}
\end{equation*}
and
\begin{equation*}
  \chi_2 \approx \frac{1}{k} + \frac{4}{(\lam k)^2} - 8\frac{2+k}{(\lam k)^3(1+k)v_0}.
\end{equation*}
For sufficiently large $k$, the second moment then converges to
\begin{equation*}
  \mean{\x^2}_\infty \approx \frac{4}{(\lam k)^2}.
\end{equation*}
In Fig.~\ref{fig:limit}(a), the scaled second moment is plotted as function of scaled speed for two values of $\lam$.

\section{Stochastic thermodynamics}

For completeness, we summarize the stochastic thermodynamics of a single active particle tightly coupled to a reaction $\bullet\to\circ$ on its surface following Ref.~\cite{zimm12}. The (Gibbs) free energy of the composite system is
\begin{equation*}
  G(n_\bullet,\x;k) = U(\x;k) + \mu_\bullet n_\bullet + \mu_\circ (n-n_\bullet)
\end{equation*}
with constant $n=n_\bullet+n_\circ$ and potential energy $U(\x;k)$ depending on a parameter $k$ that is manipulated by some external agent.

The composite system is coupled to a heat bath with constant temperature, which implies the \emph{detailed balance} condition
\begin{multline*}
  \ln\frac{\kap(\x\to\x+\lam\vec e,n_\bullet\to n_\bullet-1)}{\kap(\x+\lam\vec e\to\x,n_\bullet-1\to n_\bullet)} = -\delta G \\ = -[G(n_\bullet-1,\x+\lam\vec e)-G(n_\bullet,\x)] \\
  = \mu_\bullet-\mu_\circ - [U(\x+\lam\vec e)-U(\x)].
\end{multline*}
The right hand side is independent of the number of fuel molecules. Hence, it is sufficient to distinguish forward and backward direction and to drop $n_\bullet$ as argument leading to the \emph{local detailed balance} condition
\begin{equation*}
  \frac{\kap^+(\x)}{\kap^-(\x+\lam\vec e)} = e^{\Delta\mu-[U(\x+\lam\vec e)-U(\x)]}
\end{equation*}
with $\Delta\mu=\mu_\bullet-\mu_\circ$. If $\Delta\mu>0$ then the system can reduce its free energy through converting $\bullet\to\circ$.

The first law for a single conversion reads~\cite{seifert12}
\begin{equation}
  0 = \delta E + \delta Q_1
  \label{eq:fl:c}
\end{equation}
with internal energy $\delta E=\delta G+\delta S_\text{sol}=\delta U-\Delta\mu+\delta S_\text{sol}$, where $\delta U=U(\x+\lam\vec e)-U(\x)$ and $\delta Q_1$ is the heat exchanged with the heat bath. Strictly speaking, one should include an entropy change $\delta S_\text{sol}$ of the solvent due to the change of chemical composition~\cite{seifert11a}.

We now split changes into those due to the diffusion in the external potential (governed by $\mathcal L_0$) and due to the chemical conversions (governed by $\mathcal L_1$). The diffusive contributions obey the first law 
\begin{equation*}
  \dot{\mathcal W} = \dot{\mathcal U}_0 + \dot{\mathcal Q}_0
\end{equation*}
on the \emph{ensemble level} with work rate $\dot{\mathcal W}=\mean{\dot k\partial_kU}$ performed by an external agent. The total system obeys the first law
\begin{equation}
  \mean{\dot n}\Delta\mu + \dot{\mathcal W} = \td{}{t}\mean{U} + \underbrace{\dot{\mathcal Q}_0 + \dot{\mathcal Q}_1 + \dot{\mathcal S}_\text{sol}}_{\sig\geqslant 0}
  \label{eq:fl}
\end{equation}
with average net rate $\mean{\dot n}$ of chemical events and total entropy production rate $\sig$, which can be shown to be non-negative (\emph{e.g.}, through applying the Jensen inequality to the fluctuation theorem). We split the energy change
\begin{equation*}
  \td{}{t}\mean{U} = \dot{\mathcal U}_0 + \dot{\mathcal U}_1,
\end{equation*}
which implies
\begin{equation*}
  \dot{\mathcal Q}_1 = \mean{\dot n}\Delta\mu - \dot{\mathcal U}_1 - \dot{\mathcal S}_\text{sol}
\end{equation*}
in agreement with Eq.~\eqref{eq:fl:c}.

Integrating Eq.~\eqref{eq:fl} along a closed cycle in parameter space, we define $W_\text{in}=\oint\dd t\;P_\text{in}$ with $P_\text{in}=\mean{\dot n}\Delta\mu$ and $W_\text{out}=\oint\dd t\;\dot{\mathcal W}$. Hence,
\begin{equation*}
  W_\text{in}+W_\text{out} \geqslant \oint\dd t\;\td{}{t}\mean{U} = 0
\end{equation*}
since the average potential energy of the particle does not change for a full cycle. Clearly, the thermodynamic efficiency $\eta=-W_\text{out}/W_\text{in}\leqslant 1$ is thus bounded by one.

\section*{Numerical simulations}

For the numerical simulations, we solve the discretized Langevin equations
\begin{gather*}
  \x(t+\delta t) = \x(t) - k\x\delta t + \sqrt{2\delta t}\nois(t), \\
  \vhi(t+\delta t) = \vhi(t) + \sqrt{2\delta t}\eta(t)
\end{gather*}
corresponding to Eq.~\eqref{eq:L0} with time step $\delta t=10^{-3}$. Here, $\nois$ and $\eta$ are normal Gaussian (zero-mean and unit variance) noise. After advancing particle position and orientation, we perform a kinetic Monte Carlo simulation with rates $\kap^\pm(\x,\vhi)$ sampling the chemical events described by Eq.~\eqref{eq:L1}:

\SetInd{1em}{0em}
\begin{algorithm}[h]
  \SetAlgoNoLine
  \While{$t<\delta t$}{
    update rates $\kap^\pm$\;
    $\xi\sim U(0,\kap^++\kap^-)$\;
    \eIf{$\xi<\kap^+$}{
      $n\longleftarrow n+1$\;
      $\x\longleftarrow \x+\lam\vec e$\;
    }{
        $n\longleftarrow n-1$\;
        $\x\longleftarrow \x-\lam\vec e$\;
    }
    exponentially distributed waiting time:
    $\xi\sim U(0,1)$\;
    $t\longleftarrow t+\ln(1/\xi)/(\kap^++\kap^-)$\;
  }
  $t\longleftarrow t-\delta t$\;
  \caption{Chemical events}
\end{algorithm}

In Fig.~\ref{fig:num} we show that the numerical simulations reproduce the analytical results for the cross correlations $\chi_1=\mean{\vec e\cdot\x}$ with \begin{equation}
  \chi_1 = \frac{v_0}{1+k(1+D_\text{c})}
  \label{eq:chi1}
\end{equation}
and $\mean{\x^2}\approx2/k+v_0\chi_1/k$ using the solution Eq.~\eqref{eq:chi1}.

\begin{figure}[b!]
  \centering
  \includegraphics{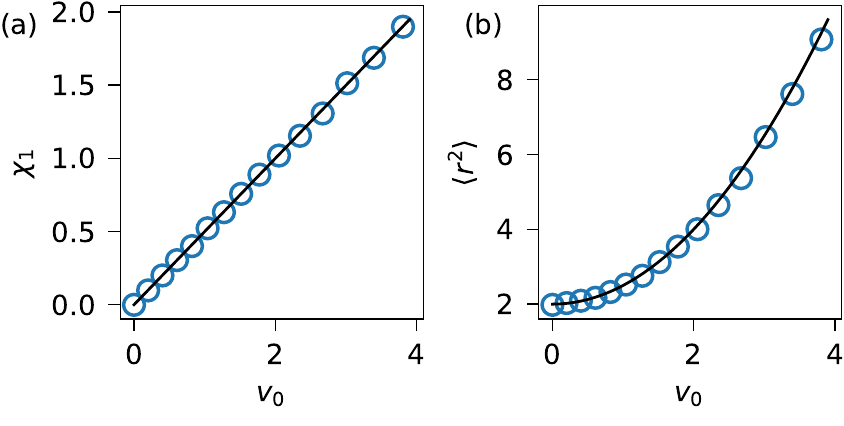}
  \caption{(a)~The cross correlations $\chi_1$ as a function of bare speed $v_0$ for $k=1$ and $D_0=\lam=10^{-3}$. Symbols are numerical results and solid line is Eq.~\eqref{eq:chi1}. (b)~The second moment $\mean{\x^2}$ for the same parameters.}
  \label{fig:num}
\end{figure}

\end{document}